\documentclass[conference]{IEEEtran}

\usepackage{cite}   
\usepackage{graphicx}  
\usepackage{amsmath, amsopn, psfrag, amsthm}   
\usepackage{amssymb}
\usepackage{color}

\usepackage{algorithm}

\usepackage{algorithmic}

\def\bfPhi{{\boldsymbol{\Phi}}}
\def\bftheta{{\boldsymbol{\theta}}}
\def\bfphi{{\boldsymbol{\phi}}}

\usepackage{psfrag}
\usepackage{epsfig}
\graphicspath{{./img/}}
\DeclareGraphicsExtensions{.pdf,.jpeg,.png, .eps}
\usepackage{amssymb,amsmath,mathtools}

\usepackage{cite}   
\usepackage{graphicx}  
\usepackage{amsmath, amsopn}   
\usepackage{amssymb}
\usepackage{color}
\usepackage{psfrag}
\usepackage{epsfig}
\usepackage{stfloats}
\usepackage{lipsum}
\usepackage{multicol}
\graphicspath{{./img/}}
\DeclareGraphicsExtensions{.pdf,.jpeg,.png}
\begin{document}
\setlength{\columnsep}{.261in}
\title{Robust Geometry-Based User Scheduling for Large MIMO Systems Under Realistic Channel Conditions}
\author{
\IEEEauthorblockN{Manijeh Bashar\IEEEauthorrefmark{1}, Alister G. Burr\IEEEauthorrefmark{1}, Dick Maryopi\IEEEauthorrefmark{1}, Katsuyuki Haneda\IEEEauthorrefmark{2}, and Kanapathippillai Cumanan\IEEEauthorrefmark{1}}

\IEEEauthorblockA{\IEEEauthorrefmark{1}Department of Electronic Engineering, University of York, Heslington, York, YO10 5DD, UK \\Email: \{mb1465, alister.burr, dm1110,kanapathippillai.cumanan\}@york.ac.uk}

\IEEEauthorblockA{\IEEEauthorrefmark{2}The Aalto
University School of Electrical Engineering, 02150 Espoo, Finland.} Email: {katsuyuki.haneda@aalto.fi}}
%
\maketitle
%
%
\begin{abstract}
The problem of user scheduling with reduced overhead of channel estimation in the uplink of Massive multiple-input multiple-output (MIMO) systems has been considered. A geometry-based stochastic channel model (GSCM), called the COST 2100 channel model has been used for realistic analysis of channels. In this paper, we propose a new user selection algorithm based on knowledge of the geometry of the service area and location of clusters, without having full channel state information (CSI) at the base station (BS). The multi-user link correlation in the GSCMs arises from the common clusters in the area. The throughput depends on the position of clusters in the GSCMs and users in the system. Simulation results show that although the BS does not require the channel information of all users, by the proposed geometry-based user scheduling algorithm the sum-rate of the system is only slightly less than the well-known greedy weight clique scheme. Finally, the robustness of the proposed algorithm to the inaccuracy of cluster localization is verified by the simulation results.
\vspace{.12cm}

{{\textbf{\textit{Keywords}:}} COST 2100 channel model, geometry-based stochastic channel models, Massive MIMO, user scheduling, zero-forcing, {cluster localization}.}
\end{abstract}
\section{Introduction}
 \let\thefootnote\relax\footnotetext{The work of A. G. Burr and K. Cumanan was supported by H2020- MSCA-RISE-2015 under grant number 690750. The work on which this paper is based was carried out in collaboration with COST Action CA15104 (IRACON). In addition, the work is supported by LPDP (Indonesia Endowment Fund for Education).}
Massive multiple-input multiple-output (MIMO) is a promising
technique to achieve high data rate \cite{massivetddMarzetta14,ouricc1,ouricc2}. However, high performance multiuser MIMO (MU-MIMO) uplink techniques rely on the availability of full channel state information (CSI) of all user terminals at the base station (BS) receiver, which presents a major challenge to their practical implementation. This paper considers an uplink multiuser system where the BS is equipped with $M$ antennas and serves $K_s$ decentralized single antenna users ($M\gg K_s$). In the uplink mode, the BS estimates the uplink channel and use a linear receivers to separate the transmitted data. The BS receiver uses the estimated channel to implement the zero-forcing (ZF) receiver which is suitable for Massive MIMO systems. To investigate the performance of MIMO systems, an accurate multi-user channel model is necessary. Most standardized MIMO channel models such as IEEE $802.11$, the 3GPP spatial model, and the COST 273 model rely on clustering \cite{standard}. Geometry-based stochastic channel models (GSCMs) consider more physical reality of clusters such as their relative locations to the BS and users in the cell to investigate the performance of MIMO systems \cite{Molish_tufvesson}. This paper investigates the throughput in the uplink for the Massive MIMO with carrier frequency around 2 GHz, but the principles can also apply to other frequency bands, including mmWave.

Most existing Massive MIMO techniques rely on the availability of the full CSI of all users at the BS, which presents a major challenge of channel estimation in implementing Massive MIMO. As a result, Massive MIMO techniques with reduced CSI requirement are of great interest. Recently, a range of user scheduling schemes have been proposed for Massive MIMO systems. Most of these, including \cite{Lee14user}, require accurate knowledge of the channel from all potential users to the BS --which in Massive MIMO case is completely infeasible to obtain; \cite{XuFDDuser} proposed a greedy user selection scheme by exploiting the instantaneous CSI of all users. However, in this paper we focus on a simplified and robust user scheduling algorithm, by considering the effect of the cell geometry.
\subsection{Contributions of This Work}
Our study on a new user selection algorithm considers high frequency stochastic geometry-based channels with large numbers of antennas at the BS receiver. Given a map of the area of the micro-cell, we perform efficient user scheduling based only on the position of users and clusters in the cell. In GSCMs, grouping multipath components (MPCs) from common clusters cause high correlation which reduces the rank of the channel \cite{Alister10ISWCS, katsumimocost, cairespatial13inftheory}. In this paper, we investigate the effect of common clusters on the Massive MIMO multi-user performance.
Our results and contributions are summarized as follows: 

\textbf{1)} We show a novel user scheduling scheme for cellular systems equipped with a large antenna array at the BS. Using the map of the area and positions of users, the new user scheduling scheme works \textit{without CSI at the BS, as far as the location of multipath clusters is known}. Assuming the positions of the clusters in the area are fixed, cluster localization can be done offline. 
\\\
{\textbf{2)}
For large numbers of transmit antennas and users, it is shown that the throughput benefits from multiuser diversity, even under the \textit{no-CSI condition}. Simulation results show significant performance improvement compared to conventional user scheduling algorithms, especially for indoor and outdoor micro-cells. The proposed scheme significantly reduces the overhead channel estimation in Massive MIMO systems.
\\\
{\textbf{3)} The robustness of the proposed algorithm to the uncertainties of cluster localization is demonstrated through numerical simulations taking into account the error bounds of the SAGE parameter estimates. }

Note that in this paper, uppercase and lowercase
boldface letters are used for matrices and vectors, respectively.
The notation $\mathbb{E}(\cdot)$ denotes expectation; $|\cdot|$ stands for the absolute value. The conjugate
transpose of vector $\textbf{x}$ is $\textbf{x}^{H}$. Moreover, $\textbf{X}^{\dag}$, $\textbf{X}^{-1}$ and $\textbf{X}^{T}$ 
denote the pseudo-inverse, inverse and transpose of matrix $\textbf{X}$, respectively. The Kronecker product of $\textbf{X}$ and $\textbf{Y}$ is presented by $\textbf{X} \otimes\textbf{Y}$. Finally, $\text{vec}(\textbf{X})$ denotes the column vector obtained by stacking the columns of the matrix $\textbf{X}$.
The rest of the paper is organized as follows. Section II describes
the system model. The proposed user scheduling scheme is presented in Section III. {The robustness of the proposed user scheduling algorithm  to cluster localization errors is investigated in Section IV.} Numerical
results are presented in Section V. Finally, Section VI concludes
the paper. 
\vspace{-.2cm}\section{SYSTEM MODEL}
Consider uplink transmission in a single cell with $M$ antennas at the BS and $K$ single antenna {mobile stations (MSs)} on the same time-frequency resource. Here, we assume TDD mode where the uplink and downlink channel are the same. 
\vspace{-.2cm}
\subsection{Uplink Training}
In this section, we investigate the problem of estimating the channel in the TDD mode. Suppose $\textbf{H}\in\mathbb{C}^{M\times K}$ represents the uplink aggregate channel matrix between the users and the BS. The channel covariance matrix $\textbf{R}\in \mathbb{C}^{MK\times MK}$ is given by
\begin{equation}
\textbf{R} = \mathbb{E}\{\tilde{\textbf{h}}\tilde{\textbf{h}}^H\},
\end{equation}
where $\tilde{\textbf{h}}=\text{vec}(\textbf{H})$. For MMSE estimation of the channel, we use a pilot sequence \cite{debbah_cor_es}, \cite{kay}. Let us assume $\bfPhi_p \in \mathbb{C}^{K\times \tau_p}$ denotes pilot matrix, where $\tau_p$ is the length of pilot sequence for each user. The received pilot signal at the BS, $\textbf{Y}\in\mathbb{C}^{{M\times \tau_p}}$, is given by
\begin{equation}
\textbf{Y}=\textbf{H}\bfPhi_p+\textbf{N},
\end{equation}
where $\text{vec}(\textbf{N})\sim \mathcal{CN}(0,\sigma_n^2\textbf{I})$ denotes circularly symmetric complex Gaussian noise, and $\textbf{I}\in\mathbb{C}^{M\tau_p\times M\tau_p}$ is the identity matrix. The Bayesian MMSE estimator of the channel is given by \cite{debbah_cor_es}
\begin{equation}
\tilde{\textbf{h}}_{MMSE}=\textbf{R}\tilde{\bfPhi}_p^H(\tilde{\bfPhi}_p\textbf{R}\tilde{\bfPhi}_p^{H}+\sigma_n^2\textbf{I})^{-1}\tilde{\textbf{y}},
\label{htilde}
\end{equation}
where $\tilde{\bfPhi}_p=\bfPhi_p^T \otimes \textbf{I}$ and $\tilde{\textbf{y}}=\text{vec}(\textbf{Y})$.
\vspace{-.2cm}
\subsection{Uplink Transmission}
The $M \times 1$ received signal at the BS when $K_s\, (K_s\ll M)$ users have been selected from the pool of $K$ users, is given by 
\begin{equation}
\textbf{r}= \sqrt{p_k}\textbf{H}_s\textbf{x}+\textbf{n}, 
\end{equation}
where $\textbf{x}$ represents the symbol vector of $K_s$ users, and is constrained to have total expected power of $ \mathbb{E}\left\{|\textbf{x}^H\textbf{x}|\right\}=K_s$, $p_k$ 
is the average uplink transmit power of the $k$th user and $\textbf{H}_s$ denotes the aggregate $M \times K_s$ channel of all selected users. 
 The BS is assumed to have CSI only of the selected users. We are interested in a linear ZF receiver which can be provided by evaluating the pseudo-inverse of the estimated channel, $\tilde{\textbf{H}}_s$, the aggregate channel of all selected users according to $\mathbf{W}=\tilde{\textbf{H}_s}^{\dag}=(\tilde{\textbf{H}_s}^{H}\tilde{\textbf{H}}_s)^{-1}\tilde{\textbf{H}_s}^{H}$ \cite{GoldsmithJurnal,myiet_master,my-master-vtc}. Then after using the detector, the received signal at the BS is
\begin{equation}
\textbf{y}= \sqrt{p_k}\textbf{W}\textbf{H}_s\textbf{x}+\textbf{W}\textbf{n}.
\label{yy}
\end{equation}
Let us consider equal power allocation between users, i.e. $p_k=\frac{P}{K_s},~\forall k $, in which $P$ denotes the total power. The achievable sum-rate of the system is obtained as
\begin{equation}
R=\sum_{k=1}^{K_s}{\log_{2}
 \bigg({{  1+\frac{p_k|{\mathbf{w}_k}{\mathbf{h}_k}|^2}{1+\sum_{i=1,i\ne{k}}^{K}p_i|\mathbf{w}_k\mathbf{h}_i|^2}}}\bigg)},
\end{equation}
where $\textbf{w}_k$ and $\textbf{h}_k$ are respectively the $k$th rows of the matrix $\textbf{W}=[\textbf{w}_1^T,\textbf{w}_2^T,\cdots,\textbf{w}_{K_s}^T]^T$, and  the \textit{k}th column of  $\textbf{H}_s=[\textbf{h}_1,\textbf{h}_2,\cdots,\textbf{h}_{K_s}]$.

If perfect CSI is available at the BS, and assuming Gaussian input, the ergodic capacity is given by
\begin{IEEEeqnarray}{rCl}
C = \mathbb{E} \left\{\log_2\det  \left(\textbf{I}+\dfrac{P}{K_s}\textbf{H}_s\textbf{H}_s^H\right)\right\},
\end{IEEEeqnarray}
where the term $\frac{P}{K_s}$ is due to the equal-power allocation and $\textbf{I}$ refers to an identity matrix.
\subsection{Geometry-based Stochastic Channel Model}
\begin{figure}
\center
\includegraphics[width=68mm]{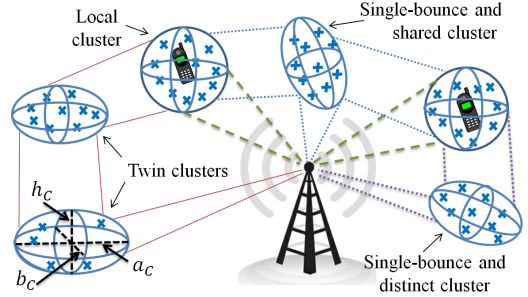}
\vspace{-0.07in}
\caption{The general description of the cluster model. The spatial spreads for $C$th cluster are given. The figure also gives an example of a shared cluster
and a distinct cluster.}
\label{cluster}
\end{figure}
In GSCMs the double directional channel impulse response is a superposition of MPCs as given by \cite{Costaction}
\begin{small}
\begin{equation}
h(t\!,\!\tau\!,\!\bfphi^{\text{BS}}\!,\!\bftheta^{\text{MS}}\!)\!=\!\sum_{j=1}^{N_C}\!\sum_{i=1}^{N_p}\!a_{i,j}\delta(\phi^{\text{BS}}-\phi_{i,j}^{\text{BS}})\delta(\theta^{\text{MS}}\!-\!\theta_{i,j}^{\text{MS}})\delta(\tau-\tau_{i,j}),\!
\label{h1}
\end{equation}
\end{small}
where $N_p$ denotes the number of MPCs, $t$ is time, $\tau$ denotes the delay, $\delta$ denotes the Dirac delta function, and $\bfphi^{\text{BS}}$ and $\bftheta^{\text{MS}}$ represent the direction of departure (DoD) and direction of arrival (DoA) respectively. 

Similar to \cite{Costaction}, we group the MPCs with similar delay and directions into clusters. The circular visibility region (VR) determines whether the cluster is active or not for a given user. The MPC's gain scales by a transition function of the VR that is given by \cite{Costaction}
\begin{equation}
A_{\text{VR}}(\bar{\boldsymbol r}_{\rm MS})=\dfrac{1}{2}-\dfrac{1}{\pi}\arctan\left(\dfrac{2\sqrt{2}\left(L_c+d_{\text{MS,VR}}-R_C\right)}{\sqrt{\lambda L_c}}\right),
\label{vr}
\end{equation}
where $\bar{\textbf{r}}_{MS}$ is a position vector, $R_C$ denotes the VR radius, $L_C$ is the size of the transition region and $d_{\text{MS,VR}}=|| r_{\text{MS}}-r_{\text{VR}}||$ denotes the distance between the mobile station (MS) and the VR centre. The cluster power attenuation is given by 
\begin{equation}
A_C=\max (\exp[-k_\tau (\tau_C-\tau_0) ], \exp[-k_\tau (\tau_B-\tau_0) ]),
\label{AC}
\end{equation}
where $k_\tau$ denotes the power decay parameter, $\tau_B$ is the cut-off delay, and $\tau_C$ refers to the delay of a cluster.
We assume Rayleigh fading for the MPCs within each cluster. Hence, the complex amplitude of the $i$th MPC in the $j$th cluster in (\ref{h1}) is given by
 \begin{equation}
a_{i,j}=L_pA_{VR}\sqrt{A_C A_{\text{MPC}}},
\label{a}
\end{equation} 
where $L_p$ is the channel path loss, $A_{\text{MPC}}$ is the Rayleigh-faded power of each MPC. For the non-line-of-sight (NLoS) case of the micro-cell scenario, the path loss is
$
L = 26 \log_{10}d_{\text{BS,MS}}+20\log_{10}(4\pi/\lambda),
$
where $d_{\text{BS,MS}}$ and $\lambda$ denote the distance from the BS to the MS and the wavelength {in meters}, respectively.
\vspace{-0.2in}

Finally, assuming a linear array response at the BS side the channel matrix is given (\ref{H3}) (defined at the top of this page), where $\textit{C}(k)$ denotes the clusters seen by the $k$th user and $\alpha=-2\pi\frac{d}{\lambda}$, where $d$ denotes the spacing between two antenna elements. Note that the index $\small{\text{BS}}$ is dropped for simplicity. In GSCMs, shared (common) clusters can reduce the rank of the channel and the capacity of the system, especially at finite signal-to-noise ratio (SNR). These common clusters also affect the multiplexing gain of the system. Fig. \ref{cluster} illustrates the concept of common and distinct clusters.
\begin{figure*}[!t]
\normalsize
\begin{IEEEeqnarray}{rCl}\begin{split}
\small
\!\textbf{H}_s\!=\!
\begin{bmatrix} 
\!\sum_{j\in \textit{C}(1)}\!\sum_{i=1}^{N_p}a_{i,j} & \!\sum_{l\in \textit{C}(2)}\!\sum_{i=1}^{N_p}a_{i,l}  & \!\ldots\! & \sum_{m\in \textit{C}(K_s)}\sum_{i=1}^{N_p}a_{i,m} \\
\!\sum_{j\in \textit{C}(1)}\!\sum_{i=1}^{N_p}a_{i,j}e^{j\alpha\sin\phi_{i,j}} & \!\sum_{l\in \textit{C}(2)}\!\sum_{i=1}^{N_p}a_{i,l}e^{j\alpha\sin\phi_{i,l}} & \ldots & \!\sum_{m\in \textit{C}(K_s)}\!\sum_{i=1}^{N_p}a_{i,m}e^{j\alpha\sin\phi_{i,m}} \\
\vdots & \vdots & \ddots & \vdots \\
\!\sum_{j\in \textit{C}(1)}\!\sum_{i=1}^{N_p}a_{i,j}e^{j\alpha(M-1)\sin\phi_{i,j}} & \!\sum_{l\in \textit{C}(2)}\!\sum_{i=1}^{N_p}a_{i,l}e^{j\alpha(M-1)\sin\phi_{i,l}} & \ldots & \sum_{m\in \textit{C}(K_s)}\!\sum_{i=1}^{N_p}a_{i,m}e^{j\alpha(M-1)\sin\phi_{i,m}}
\end{bmatrix},
\label{H3}
\end{split}
\end{IEEEeqnarray}
\vspace{-.24in}
\hrulefill
\end{figure*}
\section{Geometry-based User Scheduling}
In this section, we consider user scheduling with ZF based on the position of clusters and users in a cell. In order to avoid a huge channel estimation load in the uplink of a Massive MIMO system with many users and antennas, we propose to estimate only the channels of the selected users. The reduction in the amount of channel estimation required between each transmit and receive antenna is the important result of the proposed scheme. The gain achieved by selecting users with the strongest channel is referred to as multiuser diversity and requires CSI of all users \cite{GoldsmithJurnal}. When the number of clusters is less than the number of BS antennas and all clusters are shared between the users, it is impossible to achieve the maximum multiplexing gain \cite{Alister10ISWCS,burrijas}. 
However, we propose a new user selection scheme which relies on maximizing the number of distinct clusters seen by the scheduled users. In the next subsection, we present a scheme to select users which maximizes the long term (over time-varying channels due to movement of a user) sum rate. As it is based on the position of the users and does not need the estimated channel of all users in the uplink, it is considered to be a practical user selection scheme for large MIMO systems. 
\subsection{Proposed Geometry-based User Scheduling (GUS)}
In this section, an algorithm is proposed for increasing the system throughput based on the geometry of the system and without estimating the channels of all the users in the area. Once the set of active users has been determined, the receiver BS estimates the channels of the selected users and the users transmit data. The performance of the proposed user selection algorithm to maximize the sum-rate is evaluated. In large MIMO systems with large numbers of users estimating the channels of all users is practically difficult. So the proposed user scheduling algorithm can be an efficient way to reduce the overhead of channel estimation.\\   
First, we generate a user-cluster pathloss matrix $\textbf{V}$, as the following 
\begin{equation}
\textbf{{V}}=
\begin{bmatrix} 
v_{1}^1 & v_{2}^1 & \ldots & {v}_{N_C}^1 \\
v_{1}^2 & {v}_{2}^2 & \ldots & {v}_{N_C}^2 \\
\vdots & \vdots & \ddots & \vdots \\
v_{1}^K & v_{2}^K & \ldots & v_{N_C}^K
\end{bmatrix},
\end{equation}
where 
\begin{equation}
v_{i}^j=\sqrt{L_{p}^j}A_{VR,i}^j\sqrt{A_{C,i}^j},
\label{a}
\end{equation} 
where $L_{p}^j$ denotes the channel path loss for user $j$, $A_{VR,i}^j$ is the MPC power attenuation as a function of the distance between user $j$ and the centre of the VR activating the $i$th cluster and is given by (\ref{vr}); $A_{C,i}^j$ denotes the cluster power attenuation, given (\ref{AC}), by for the user $j$ and the $i$th cluster. So, the matrix $\textbf{V}$ is a function of the distance from the BS to users, the distance of the BS from clusters and from users to the centre of the VR.
The BS uses the functions $f_1(\textbf{v}_i)$ and $f_2(\textbf{v}_i)$ where $\textbf{v}_i$ is the $i$th row of matrix ${\textbf{V}}$ and we define the functions $f_1(\textbf{v})$ and $f_2(\textbf{v})$ in Algorithm 1.
\begin{algorithm}[t!]
Step 1) Initialization:~~$\mathcal{W}_{0}=[1,\cdots,K]$, $\mathcal{S}_{0}=\emptyset$, $i=1$,\caption{{\small Geometry-based User Scheduling (GUS) Algorithm}}

Step 2) Load position of users, for example by means of GPS,\\
Step 3) Generate matrix $\textbf{V}$,\\
Step 4) Greedy Algorithm:
\begin{itemize}
\item{4.1}\begin{small}
$\pi(1)=\operatornamewithlimits{argmax}\limits_{k \in \mathcal{W}_{0}}f_1(||\textbf{v}_k||)=\operatornamewithlimits{argmax}\limits_{k \in \mathcal{W}_{0}}||\textbf{v}_k||$,\\ $\mathcal{S}_{0}\gets \mathcal{S}_{0}\cup \{k\}$, $\hat{\textbf{v}}_{(i)}=\textbf{v}_{(\pi(i))}$,\\
\end{small}
\item{4.2}~If $|\mathcal{S}_{0}|<K_s$, $\mathcal{W}_{i}=\{k \in \mathcal{W}_{i-1}, k \ne \pi(i) \}$,\\
\item{4.3}
\begin{small}
$\pi(i)=\operatornamewithlimits{argmin}\limits_{k \in \mathcal{W}_{i-1}}f_2(\textbf{v}_k,\hat{\textbf{v}}_{(i)})=\operatornamewithlimits{argmin}\limits_{k \in \mathcal{W}_{i-1}}\frac{  |\textbf{v}_k\hat{\textbf{v}}_{(i)}^{*}|  }{  ||\textbf{v}_k||||\hat{\textbf{v}}_{(i)}||  } \}$,~$\mathcal{S}_{0}\gets \mathcal{S}_{0}\cup \{k\}$, $\hat{\textbf{v}}_{(i)}=\textbf{v}_{(\pi(i))}$,
\end{small}
\item{4.4}~then $i=i+1$, and go to step 4.3, Else, end.
\end{itemize}
Step 5) The BS estimates the channels of the selected users.
\end{algorithm}
{Suppose $\mathcal{W}_{0}$ contains user indices considered in the proposed algorithm. The proposed algorithm, executed at each symbol time, using the position of users as described in the above algorithm, always selects  $K_s$ users. Finally, $\mathcal{S}_{0}$ contains $K_s=|\mathcal{S}_{0}|$ indices of the selected users.}  
As described in step $4.1$ in Algorithm 1, the algorithm starts by calculating the summation over all cluster powers, i.e. $f_1(\parallel\textbf{v}_k\parallel|) = \parallel\textbf{v}_k\parallel,~ \forall ~k$, and selects the user with the strongest received power at the BS. Then in the next step, the proposed algorithm finds a set of users with smallest orthogonality to the selected users. Here, orthogonality among the user $k$ and the user $j$ is defined as $f_2(\textbf{v}_k,{\textbf{v}}_{j})=\frac{ |\textbf{v}_k{\textbf{v}}_{j}^{*}| }{ ||\textbf{v}_k||||{\textbf{v}}_{j}||}$. Note that MPCs from shared clusters cause high correlation which reduces the rank of the channel. Hence, the proposed Algorithm 1 selects users with lowest correlation to improve the throughput. The capacity analysis have have been investigated in \cite{ouriet_mic}. Moreover, investigating the effectiveness of the proposed user scheduling scheme in the distributed Massive MIMO systems \cite{ourjournal1,ourjournal2} is an interesting topic for future work.
\vspace{-.1cm}
\section{Robustness of the Proposed User Scheduling Algorithm}
\subsection{Cluster Localization}
 The BS can estimate the direction of arrival \cite{handbook}, and hence the direction of the scattering objects should be available at the BS.
There is a well-known algorithm to estimate the delay, DoA and the DoD of the channel paths; SAGE-based algorithm \cite{crlbc, crlbj}. As a result, the BS can identify the direction of the clusters which can be seen by the users in the cell area, and hence build up a map of the location of the scattering objects. The convenient tool that has overcame the challenge of making the position of the scatterers available is the use of environment maps \cite[Chapter~2]{Costaction}, which also shows how measured DoA can be identified with physical objects in the environment, and hence can be located on the map. Successive interference cancellation has also been introduced in \cite{molish-tuf-position} for scattering object identification: it uses the channel impulse response peaks in the delay domain to map scatterers to two-dimensional coordinates.
\vspace{-.1cm}
\subsection{{Robustness}}
In order to study the robustness of the proposed algorithm to possible uncertainty in cluster localization, we assume the well-known SAGE algorithm \cite{crlbc,crlbj} as a means to estimate DoAs and delays at the BS, operating offline, as mentioned above. In cluster localization, we consider a receiver BS with an $M$-element antenna array located at an reference point \cite{crlbc}-\cite{crlbj}. Moreover, we consider planar wavefronts with $M_x$ sensor at each direction.
The closed-form Cramer-Rao lower bound (CRLB) for the delay, azimuth ($\phi$) and elevation ($\theta$) of the path are given by \cite{crlbc}
\begin{subequations}
\begin{eqnarray}
& \text{CRLB}(\tau) =\dfrac{1}{\gamma_O}\dfrac{1}{8\pi^2BW} \label{crlb1}\\
& \text{CRLB}(\theta) = \dfrac{1}{\gamma_O}\dfrac{M}{2\Delta \cos (\theta)} \label{crlb2}\\
& \text{CRLB}(\phi) = \dfrac{1}{\gamma_O}\dfrac{M}{2\Delta}\label{crlb3},
\end{eqnarray}
\label{crlb}
\end{subequations}
where $BW$ is the bandwidth, $\Delta=4\pi^2(\frac{d}{\lambda})^2(\frac{7}{3}M_x^3-8M_x^2+\frac{29}{3}M_x-4)$, $\gamma_O=MIN_c|f(\phi)|^2\gamma_I$, where $I$ is the number of periods of the received signal, $N_c$ denotes the length of the used pseudonoise
(PN) sounding sequence available at the receiver and $\gamma_I$ is the SNR at the input of each antenna \cite{crlbc}-\cite{crlbj}. Moreover, the antenna electric field pattern can be given by
$
f(\phi)=0.67+2.67\phi-6.79\phi^2+5.7\phi^3-1.71\phi^3.
$
\\
\textit{\textbf{Remark 1:}}
The distance between the BS and single-bounce cluster ($d_{BS,C}$) is given by geometrical calculation:
\begin{IEEEeqnarray}{rCl}
(c_0\tau-d_{BS,C})^2 & = &(h_{BS}-h_{MS}+d_{BS,C}\sin(\phi))^2+ \nonumber \\ &  
&(d_{BS,MS}-d_{BS,C}\cos(\phi)\cos(\theta))^2,
\label{dis}
\end{IEEEeqnarray}
where $c_0$ denotes the velocity of light, $d_{BS,MS}$ is the distance between the user and the BS in $x-y$ plane. The distance between the user and a single-bounce cluster is easily given by $d_{MS,C}+d_{BS,C}=c_0\tau$. 
\\\\
Hence, using Remark 1, after the offline localization, the BS can build up the matrix $\tilde{\textbf{V}}$ at the beginning of each time-slot, as the following 
\begin{equation}
{\tilde{\textbf{V}}}=
\begin{bmatrix} 
\tilde{v}_{1}^1 & \tilde{v}_{2}^1 & \ldots & \tilde{v}_{N_C}^1 \\
\tilde{v}_{1}^2 & \tilde{v}_{2}^2 & \ldots & \tilde{v}_{N_C}^2 \\
\vdots & \vdots & \ddots & \vdots \\
\tilde{v}_{1}^K & \tilde{v}_{2}^K & \ldots & \tilde{v}_{N_C}^K
\end{bmatrix},
\label{vtilde}
\end{equation}
where
\begin{equation}
\tilde{v}_{i}^j=\sqrt{L_{p}^j}\tilde{A}_{VR,i}^j\sqrt{\tilde{A}_{C,i}^j},
\label{a}
\end{equation} 
where $\tilde{A}_{VR,i}^j$ and $\tilde{A}_{C,i}^j$ can be calculated by the distances obtained in (\ref{dis}). Finally, for the matrix \textbf{V}, the following equation holds
\begin{equation}
\textbf{V} = \tilde{\textbf{V}}+\textbf{E},
\label{err}
\end{equation}
where $\textbf{E}$ is due to the estimation error in cluster localization. Then, we use $\tilde{\textbf{V}}$ instead of $\textbf{V}$ in the proposed algorithm. The numerical results verify the robustness of the proposed algorithm to this error. 
\\
\textit{\textbf{Remark 2:}}
Similar to \cite{SUSGoldsmithGlobcom}, we only serve the selected users and the users which are not selected by the proposed user scheduling scheme are served in different time and channel transmission resources. Evaluation of resource assignment provides possible
directions for future work.
\section{Numerical Results and Discussion}
In this section, simulation results have been provided to validate the performance of the proposed schemes with different system parameters.
We evaluate the throughput of the system, averaging over 300 random realizations of the locations of the users, clusters and shadow fading.
A square cell with a side length of $2\times R$ has been considered; we call $R$ the cell size and also assume users are uniformly distributed in the cell. As in \cite{MarzettaMRC13}, we assume that there is no user closer than $R_{th}=0.1\times R$ to the BS. We simulate a micro-cell environment for the NLoS case and set the operating frequency $f_C=2$ GHz. The external parameters and stochastic parameters are extracted from chapter 6 of \cite{corria} and chapter 3 of \cite{Costaction}. The BS and user heights are assumed to be $h_{BS}=5$ m and $h_{MS}=1.5$ m, respectively. The number of clusters and their visibility and transition regions, specified in Section II-C, are set $N_C=3$, $R_C=50$, and $L_C=20$. Moreover, we consider $N_{P}=6$ MPCs per cluster.
The noise power is given by
$
P_n = \text{BW}  k_B  T_0  W
$,
where $\text{BW}=20$ MHz denotes the bandwidth, $k_B = 1.381 \times 10^{-23}$ represents the Boltzmann constant, and $T_0 = 290$ Kelvin denotes the noise temperature \cite{marzetta_free16}. Moreover, $W=9$ dB is the noise figure. For the sake of simplicity and without loss of generality, equal power allocation between users is assumed, ie., $p_k=\frac{P}{K_s},~\forall k$, as it is given in (\ref{yy}).
\begin{figure}[t!]
\center
\includegraphics[width=90mm]{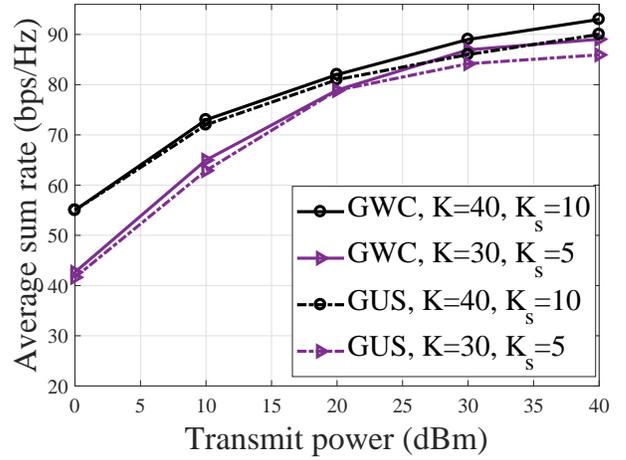}
\vspace{-.19in}
\caption{The average sum-rate vs. total transmit power for $M=200$ and different values of $K_s=10$, $K_s=5$, and $R=1000$ m.}
\label{vsp}
\end{figure}
\begin{figure}[t!]
\center
\includegraphics[width=90mm]{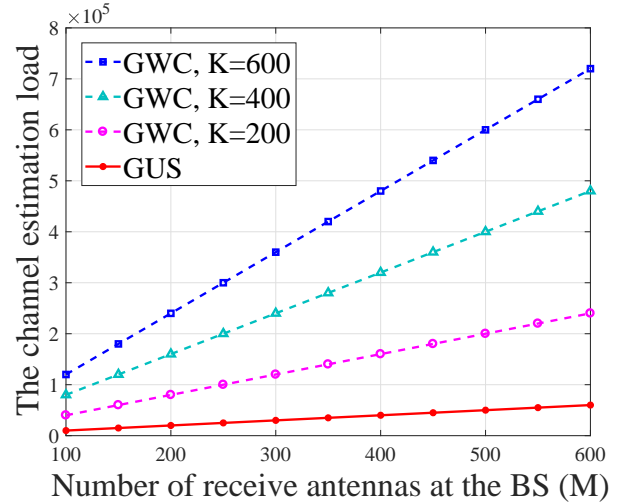}
\vspace{-.19in}
\caption{The channel estimation load vs. value of error of antennas at the receiver BS for different values of total number of users in the cell with $R=1000$ m.}
\label{vsM}
\end{figure}
\begin{figure}[t!]
\center
\includegraphics[width=90mm]{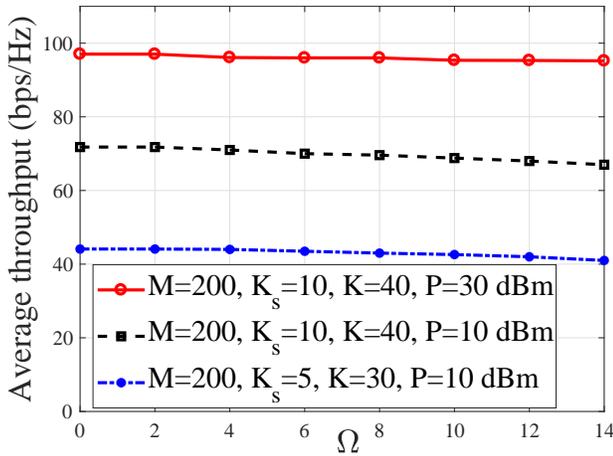}
\vspace{-.19in}
\caption{{The average sum-rate vs. the estimation error for different values of total number of selected users in the cell and the cell size.}}
\label{e}
\end{figure} 
\subsection{Numerical Results}
For this network setup, the average sum-rate is evaluated for the three scenarios. In the GUS scheme, it has been proposed that the receiver BS selects users that maximize the number of distinct clusters in the cell. We evaluate the average throughput of the greedy weight clique
(GWC) scheme \cite{SUSGoldsmithGlobcom, ITC09_Userselection_GWC}. For the case of GWC, similar to \cite{ITC09_Userselection_GWC}, we set the optimal channel direction
constraint to achieve the best performance for GWC, so the complexity of GWC is much higher than GUS.

Fig. \ref{vsp} depicts the average sum-rate with total number of receive antennas at the BS $M=200$, and two values of the number of selected users $K_s=10$ and $K_s=5$ while adopting the proposed scheme with ZF receiver. As expected, since GWC exploits perfect CSI, it has the best throughput. 
The amount of channel estimation load required in both GWC and the proposed GUS is presented in Fig. \ref{vsM}, where we use $N^{load}=2MK_s$ to calculate the total channel estimation load.
As the figure shows the channel estimation load of the proposed GUS is far less than that of the GWC scheme.

{To investigate the robustness of the proposed scheme to different values of the error, we set $|e|=\Omega\times \sqrt{\text{CRLB}(\rho)}$, where $|e|$ denotes the absolute value of the estimation error, $\Omega$ is an integer number and $\text{CRLB}(\rho)$ is given by (\ref{crlb1})-(\ref{crlb3}), where the parameter $\rho$ can be the delay, azimuth and elevation. Fig. \ref{e} shows the average sum-rate with total number of receive antennas at the BS $M=400$, and two values of the number of selected users $K_s=10$ and $K_s=5$ versus $\Omega$. We set the SNR at the input of each antenna $\gamma_I=20$ dB and the bandwidth $BW=20$ MHz. Moreover, in equations (\ref{crlb1}) to (\ref{crlb3}), $M_x=5$, $N_c=127$, which are extracted from \cite{crlbc}. The figure shows the robustness of the proposed algorithm to poor cluster localization.}
\section{Conclusions}
In this paper, we have investigated GUS under uplink Massive MIMO conditions. By applying knowledge the geometry of the system (the location of clusters and the users), we suppose that the BS does not need to estimate the channels of all users and selects users based only on the location of users and clusters in the area. The results show that while sum-rate slightly decreases along with the reduced overhead of channel estimation, the proposed algorithm can be an efficient scheme to reduce the complexity of user scheduling in Massive MIMO systems. {The proposed algorithm shows good robustness against the estimation error of cluster locations.}
\bibliographystyle{IEEEtran}
\bibliography{ewarxive} 
\end{document}